\documentclass[journal,10pt,oneside]{IEEEtran}
\usepackage{stfloats}
\usepackage{bbm}
\usepackage{amsfonts}
\usepackage{}
\usepackage{color}
\usepackage{mathrsfs}
\usepackage{algorithm}
\usepackage{algorithmic}

\usepackage{multirow}
\usepackage{graphicx,cite,epsfig,amssymb,amsmath}
\allowdisplaybreaks[4]

\renewenvironment{cases}{\left\{\begin{array}[c]{ll}}{\end{array}\right.}

\newtheorem{lemma}{Lemma}

\graphicspath{{figure/}}  

\begin{document}
\title{Generalized Orthogonal Chirp Division Multiplexing in Doubly Selective Channels}

\author{\IEEEauthorblockN{Yun Liu, Hao Zhao, Huazhen Yao, Zeng Hu, Yinming Cui, Dehuan Wan}

\thanks{
	
	Yun Liu and Dehuan Wan are with the School of Internet Finance and Information Engineering, Guangdong University of Finance,  Guangzhou 510521, China (e-mail: yunliu@gduf.edu.cn, wan\_e@gduf.edu.cn). Hao Zhao is with the Department of Communication Engineering, Guangzhou Maritime University, Guangzhou 510725, China (e-mail:zhaohao@gzmtu.edu.cn). Yuazhen Yao and Zeng Hu are with the School of Information Science and Technology, Zhongkai University of Agriculture and Engineering, Guangzhou 510225, China (e-mail:11685450@qq.com, huzeng@zhku.edu.cn). Yinming Cui is with the School of Electronics and Information Engineering, South China University of Technology, Guangzhou 510640, China (e-mail: ymcui@scut.edu.cn).}	
}

\maketitle
\begin{abstract}
In recent years, orthogonal chirp division modulation (OCDM) has gained attention as a robust communication waveform due to its strong resistance to both time-domain and frequency-domain interference. However, similar to orthogonal frequency division multiplexing (OFDM), OCDM suffers from a high peak-to-average power ratio (PAPR), resulting in increased hardware costs and reduced energy efficiency of the transmitter's power amplifiers. In this work, we introduce a novel unitary transform called the Generalized Discrete Fresnel Transform (GDFnT) and propose a new waveform based on this transform, named Generalized Orthogonal Chirp Division Modulation (GOCDM). In GOCDM, data symbols from the constellation diagram are independently placed in the Generalized Fresnel (GF) domain. We derive the GF-domain channel matrix for the GOCDM system under time-frequency doubly selective channels and leverages the sparsity of the GF-domain channel matrix to design an iterative receiver based on the message-passing algorithm. Simulation results demonstrate that GOCDM achieves better PAPR performance than OCDM without compromising bit error rate (BER) performance.
\end{abstract}

\begin{IEEEkeywords}
	Generalized Discrete Fresnel Transform (GDFnT), Generalized Orthogonal chirp division multiplexing (GOCDM), time-varying, vehicular communications, underwater acoustic communications,  message passing.  
\end{IEEEkeywords}

\IEEEpeerreviewmaketitle

\section{Introduction}
In wireless communication channels, the signal reaches the receiver via multipath propagation. The differences in time delays among various channel paths, described by delay spread, induce frequency-selective fading in the signal. Additionally, the relative motion between transceivers induces Doppler shifts in the signal copies along each path. The Doppler shifts may vary across different paths. These variations are usually characterized by the channel's Doppler spread. The Doppler effect causes the signal to undergo time-selective fading. Hence, in highly mobile wireless channels, the signal undergoes time-and-frequency-selective fading, commonly known as doubly selective (DS) fading \cite{TseBook}. 

Orthogonal Frequency Division Multiplexing (OFDM) is a widely used waveform in wireless communications. Thanks to its ability to achieve spectral-efficient communication over frequency-selective channels by simply using single-tap per-subcarrier equalization, OFDM has been adopted in numerous wireless communication systems over the past decades, such as wireless local area networks (WLAN) \cite{WLAN}, the 3rd Generation Partnership Project Long-Term Evolution (3GPP LTE)  wireless broadband communication systems \cite{LTE}, the fifth generation of mobile network (5G) \cite{5G}, and others. However, the orthogonality of subcarriers in OFDM signals is highly vulnerable to disruption by channel Doppler shifts, resulting in inter-carrier interference (ICI) \cite{TCE_ICI}. The presence of ICI greatly increases the overhead of channel estimation and symbol detection in OFDM, including the need for more pilots and higher computational complexity \cite{ICI_Haggman,DS_OFDM_ChnnlEst,ICI_Wen}. Additionally, OFDM cannot exploit the channel's frequency diversity with uncoded transmission, as each data symbol is conveyed on a single subcarrier. Furthermore, OFDM signals exhibit an excessively high peak-to-average power ratio (PAPR) \cite{TCE_OFDM-PAPR,TCE_OFDM-PAPR2}. To avoid nonlinear distortion of the transmitted signal, the power amplifier must operate with a larger dynamic range, which negatively impacts both the cost and energy efficiency of the transmitter \cite{PAPR_OFDM_Analyze,TCE_PAPR}. 

For systems where the transmitter is sensitive to power and energy efficiency, or those seeking to achieve higher transmission power with a given power amplifier, discrete Fourier transform (DFT) precoded OFDM (DFT-OFDM), also known as single carrier (SC) block transmission, is an  attractive waveform  due to its very low PAPR \cite{TCE_SC-FDE}. For instance, it has been adopted by the LTE standard as the uplink transmission technology for mobile communication systems \cite{LTE, SC_TaoJun}. Additionally, since each data symbol in DFT-OFDM distributes its energy across the entire frequency domain of the system, it can effectively exploit the channel's frequency diversity even with uncoded transmission. 

In \cite{OCDM_ZhaoJian}, A novel waveform named Orthogonal Chirps Division Multiplexing (OCDM) was initially proposed for fiber-optical communications. In recent years, research on OCDM has extended to RF communication \cite{OCDM_Analysis_Xiaoli_Ma,OCDM_MIMO_ChnnlEst_2023,DS_pilot_WCL_YiyinWang_2023}, underwater acoustic communication \cite{OCDM_UWA_IOT_2023, OCDM_UWA_shallowWater_ACCESS2022}, and integrated sensing and communication systems \cite{OCDM_Radar_SPL_2022, OCDM_Radar_ChnnlEst_2023}. Using an inverse discrete Fresnel transform (IDFnT), OCDM modulates a group of data symbols onto mutually orthogonal chirps that are superimposed in the time domain. Thanks to each chirp experiencing the entire duration and bandwidth of OCDM symbols, OCDM demonstrates robust interference resistance (against narrow-band noise or burst noise), as well as good capability in exploiting channel time diversity and frequency diversity \cite{OCDM_Analysis_Xiaoli_Ma}. In a quasi-static frequency selective channel, assuming sufficient guard intervals and the same spectral efficiency, OCDM and DFT-OFDM have the same performance and are better than OFDM in terms of bit error rate (BER). But when the the guard interval is insufficient, the OCDM out performs DFT-OFDM \cite{OCDM_ZhaoJian}. It has been proven that, in terms of the achievable rate, OCDM and DFT-OFDM are the optimal waveforms for frequency-selective channels, while OCDM and OFDM are optimal for time-selective channels, under the following assumptions: 1) channel state information (CSI) is known only by the receiver, and 2) the receiver performs sufficient iterative detection using decision feedback information \cite{OCDM_optimal_waveform_CL_2018,OCDM_MMSE_PIC_Bomfin_2018}. 

Since OFDM and SC allocate the energy of each data symbol to a single subcarrier and a single time slot, respectively, they are unable to fully exploit the time and frequency domain diversity of doubly selective channels. In contrast, OCDM leverages the diversity in both time and frequency domains by spreading the energy of each Chirp signal across the entire time and frequency spectrum. However, similar to OFDM, OCDM signals also exhibit a high PAPR \cite{OCDM_PAPR_AI_Ma}.

In this paper, we propose a new waveform called Generalized OCDM (GOCDM) to reduce the signal's PAPR. We then study the demodulation method of the GOCDM signal based on the DS channel with multiple lags and multiple Dopplers (MLMD). An MLMD channel comprises multiple paths, each characterized by its own path gain, time delay, and Doppler shift. The MLMD model has been widely used to describe scenarios involving relative motion between transceivers, such as vehicular communications and underwater acoustic UUV communications. \cite{OCDM_MP, MLMD_Vehicle, MLMD_highspeedTrain, MLMD_UWA}. To evaluate the proposed GOCDM detectors, we adopt two MLMD channels in the simulation. The first is an under-spread radio channel, where the coherence time is much larger than the delay spread. The second is an over-spread underwater acoustic channel, where the coherence time is comparable to or even shorter than the delay spread.
       
    The main contributions of this paper are:
    \begin{itemize}
    	\item We propose a novel unitary transform called the Generalized Discrete Fresnel Transform (GDFnT), with the conventional Discrete Fresnel Transform (DFnT) as a special case of this broader framework. Using the GDFnT, we introduce an innovative waveform named GOCDM and present a low-complexity implementation method for it.
    	
    	\item Given the gains, delays, and Doppler shifts of the channel paths, we derive the equivalent channel matrix in the generalized-Fresnel (GF) domain. Furthermore, we propose a low-complexity method to approximate this matrix as a sparse matrix. This sparse representation can be utilized to design a complexity-reduced receiver.
    	
    	\item Using the approximated GF-domain channel matrix, we describe the input-output relationship of GOCDM with a factor graph. We then propose an message passing (MP) based receiver to iteratively detect the data symbols.
    	
    	\item We evaluate the PAPR and BER performance of GOCDM in comparison to OCDM using Monte Carlo simulations. For BER assessment, we utilize both Minimum Mean Square Error (MMSE) equalization-based receivers and message passing (MP) detectors.
    \end{itemize}

The remainder of this paper is organized as follows: Section II introduces the GDFnT to support GOCDM. Section III presents the proposed GOCDM. In Section IV, the mathematical model of the GOCDM system under an MLMD channel is studied. Section V discusses the message-passing-based detector for GOCDM. Section VI provides simulations to evaluate the performance of the GOCDM system under DS channels. Finally, Section VII concludes the paper.
    
    \textit{Notation}:  Bold upright uppercase letters are used to denote matrices (e.g., $\mathbf{A}$), while bold italic lowercase letters denote vectors (e.g., $\boldsymbol{a}$). Functions with continuous and discrete variables are represented by $x(\cdot)$ and $x[\cdot]$, respectively. Some of the mathematical notations are listed as follows.

    \begin{table}[hp] \normalsize 
    	\begin{tabular}{ll} 
    		$j$ & $\sqrt{-1}$ \\
    		$ \bar p(\cdot)$ &  probability of an event\\
    	    $\mathbb{E}(\cdot)$ & expectation of a random variable \\
    	     $ (\cdot)^{H} $ & Hermitian transpose of a matrix\\
    	     $ (\cdot)^{T} $ & transpose of a matrix \\
    	     $ (\cdot)^{-1} $ & inversion of a matrix \\
    	     $ (\cdot)^{*} $ & conjugate of a complex variable \\
    	     $\mathrm{diag}(\cdot) $ & diagonal matrix converted from a vector\\
    	     $\mathcal{Z} $ & the set of integer \\
    	     $\mathcal{R} $ & the set of real numbers \\
    	     $\mathbf{I}_n$ & the $ N\times N$ identity matrix \\
    	     $\left[ \mathbf{A} \right] _{m,n}$ & the $(m,n)$-th element of matrix $\mathbf{A}$ \\
    	     $\left[ \boldsymbol{a} \right] _{m}$ & the $m$-th element of vector $\boldsymbol{a}$ \\
    	     $\Re(\cdot)$ & the real part of a complex number \\
    	     $\Im(\cdot)$ & the imaginary part of a complex number \\
    	     $\delta  [\cdot]$ & the Dirac delta function \\
    	     $ \lfloor \cdot \rfloor$ & the largest integer not greater the given number\\
    	     $\left< n \right> _N$ & $n$ modulo $N$
    	\end{tabular}
    \end{table}
 
\section{Generalized Discrete Fresnel Transform}    
Given a complex column vector $\boldsymbol{a}$ of length $MN$, where  $M$ and $N$ are positive integers, its GDFnT parameterized by $(M, N)$ is $\boldsymbol{\alpha }=\mathbf{\Theta }_{M,N}\boldsymbol{a}$, with the GDFnT transform matrix  $\mathbf{\Theta }_{M,N}$ defined as:
\begin{equation}
	\mathbf{\Theta }_{M,N} = \mathbf{\Phi }_N\otimes \mathbf{I}_M,  \label{eqn:ThetaDef}
\end{equation}
where  $\otimes$ is the Kronecker product operator, $\mathbf{\Phi }_N$ is the conventional $N$-dimensional DFnT transform matrix. The $(n,n^\prime)$th element of $ \mathbf{\Phi }_N$ is
\begin{equation} \label{Phi_n_n}
	\left[ \mathbf{\Phi }_N \right] _{n,n^{\prime}}=\frac{1}{\sqrt{N}}e^{-j\frac{\pi}{4}}e^{j\frac{\pi}{N}\left( n^{\prime}-n \right) ^2}, 
\end{equation}
$n,n^{\prime}=0,1,\cdots ,N-1$, when $N$ is even. When $N$ is odd, the value of $\left[ \mathbf{\Phi }_N \right] _{n,n^{\prime}}$ is given by
\begin{equation}
	\left[ \mathbf{\Phi }_N \right] _{n,n^{\prime}}=\frac{1}{\sqrt{N}}e^{-j\frac{\pi}{4}}e^{j\frac{\pi}{N}\left( n^{\prime}-n+\frac{1}{2} \right) ^2}.
\end{equation}
According to the definition of the Kronecker product, $\mathbf{\Theta }_{M,N}$  can be expressed  in the form of a block matrix, as shown in (\ref{blockMat_theta}) at the top of next page. 
\begin{figure*}[!ht]
	\centering
	\begin{equation}
		\mathbf{\Theta }_{M,N}=\left[ \begin{matrix}
			\left[ \mathbf{\Phi }_N \right] _{0,0}\mathbf{I}_M&		\left[ \mathbf{\Phi }_N \right] _{0,1}\mathbf{I}_M&		\cdots&		\left[ \mathbf{\Phi }_N \right] _{0,N-1}\mathbf{I}_M\\
			\left[ \mathbf{\Phi }_N \right] _{1,0}\mathbf{I}_M&		\left[ \mathbf{\Phi }_N \right] _{1,1}\mathbf{I}_M&		\cdots&		\left[ \mathbf{\Phi }_N \right] _{1,N-1}\mathbf{I}_M\\
			\vdots&		\vdots&		\ddots&		\vdots\\
			\left[ \mathbf{\Phi }_N \right] _{N-1,0}\mathbf{I}_M&		\left[ \mathbf{\Phi }_N \right] _{N-1,1}\mathbf{I}_M&		\cdots&		\left[ \mathbf{\Phi }_N \right] _{N-1,N-1}\mathbf{I}_M\\
		\end{matrix} \right]  \label{blockMat_theta}
	\end{equation}
\end{figure*}

Additionally, the GDFnT is an unitary transform, namely, the inverse generalized discrete Fresnel transform (IGDFnT) matrix $\mathbf{\Theta }_{M,N}^{-1}=\mathbf{\Theta }_{M,N}^{H}$.

\textbf{Proof.} Thanks to the properties of the Kronecker product and the traditional DFnT \cite{OCDM_ZhaoJian, matAnalysis_Meyer}, the inverse matrix of $\mathbf{\Theta }_{M,N}$ can be expressed as
\begin{align}
	\mathbf{\Theta }_{M,N}^{-1}&=\mathbf{\Phi }_{N}^{-1}\otimes \mathbf{I}_M \nonumber \\
	&=\mathbf{\Phi }_{N}^{H}\otimes \mathbf{I}_M  \nonumber \\
	&=\left( \mathbf{\Phi }_N\otimes \mathbf{I}_M \right) ^H \nonumber \\
	&=\mathbf{\Theta }_{M,N}^{H}. \label{eqn:ThetaInv}
\end{align}

\section{System Model}
\subsection{GOCDM Modulation}
\begin{figure*}[!ht]
	\includegraphics[width=18cm]{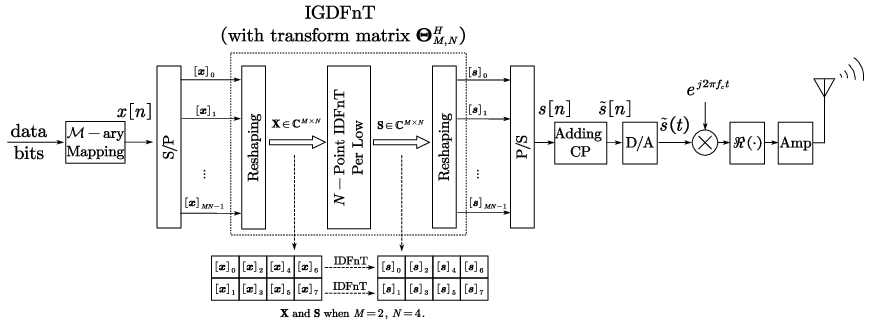}
	\caption{\small The block diagram of GOCDM transmitter.}
	\label{diagram_GOCDM}
\end{figure*}

GOCDM conveys data symbols block by block. Without loss of generality, we use a signal block as an example to introduce the principles of GOCDM. As illustrated in Fig. \ref{diagram_GOCDM}, initially, a set of independent data bits with a total length of $MN\log _2\mathcal{M}$ is mapped to $MN$ independent data symbols using an $\mathcal{M}$-ary quadrature amplitude modulation (QAM) or phase shift keying (PSK) constellation $\mathcal{X}=\{\alpha_0, \alpha_1,\dots,\alpha_{\mathcal{M}-1} \}$, where $M$, $N$, and $\mathcal{M}$ are positive integers. Let the resulting data-symbol sequence be denoted as $x_n$, $n=0$, $1$, $\dots$, $MN-1$. Through a serial-to-parallel converter, the symbol sequence is converted into the vector $\boldsymbol{x}$ with $\left[ \boldsymbol{x} \right] _n=x\left[ n \right] $. Then,  $\boldsymbol{x}$ is transformed into the vector $\boldsymbol{s}$ using an IGDFnT with transform matrix $\mathbf{\Theta }_{M,N}^{H}$ as
\begin{equation}
	\boldsymbol{s}=\mathbf{\Theta }_{M,N}^{H}\boldsymbol{x}.
\end{equation}
 Next, through a serial-to-parallel converter, the signal vector $\boldsymbol{x}$ is transformed into the signal sequence $s[n]$, $n=0,1,\dots,MN-1$, with $s\left[ n \right] =\left[ \boldsymbol{s} \right] _n
$. By adding a cyclic prefix of length $G$ to the sequence $x[n]$, we obtain the sequence $\tilde{s}[n]$ expressed as
\begin{equation}
\tilde{s}\left[ n \right] =\begin{cases}
	s\left[ n \right] ,&		0\leqslant n\leqslant MN-1\\
	s\left[ n+MN \right] ,&		-G\leqslant n<0\\
\end{cases}.
\end{equation}
After that, through a digital-to-analog converter, the discrete signal $\tilde{s}[n]$ is transformed into the continuous-time signal $\tilde{s}(t)$, which is the equivalent baseband signal of the transmitted signal. The relationship between $\tilde{s}[n]$ and $\tilde{s}(t)$ can be described as
\begin{align}
	\tilde{s}\left[ n \right] &=\sqrt{T_s}\left. \tilde{s}\left( t \right) \right|_{t=n T_s} \nonumber \\
	&= \sqrt{T_s}\tilde{s}\left( n T_s \right), -G \leqslant n\leqslant MN-1, \label{tilde_s}
\end{align}
where $T_s = T/MN$ represents the the sampling interval of the signal under ideal sampling conditions, with $T$ being the duration of a GOCDM block excluding the CP. The coefficient $\sqrt{T_s}$ in (\ref{tilde_s}) ensures that the discrete-time signal $\tilde{s}\left[ n \right]$ has the same block energy as the continuous-time one $\tilde{s}(t)$. Namely, 
\begin{equation}
	\sum_{n=-G}^{MN-1}{\left| \tilde{s}\left[ n \right] \right|^2=\int_{-T_G}^T{\left| \tilde{s}\left( t \right) \right|^2dt}}
\end{equation}
where $T_g=G T_s$ is the CP duration of the GOCDM symbol block. 
Finally, $\tilde{s}(t)$ is sent to the high-frequency module, which modulates the equivalent baseband signal onto the carrier, amplifies the power, and transmits it.

To reduce computational complexity required by $\mathbf{\Theta }_{M,N}^{H}\boldsymbol{x}$, we utilize the method illustrated in Figure  \ref{diagram_GOCDM} to equivalently implement the IGDFnT transform. The specific steps are as follows: 1) Reshape $\boldsymbol{x} $  into an $M$-row, $N$-column matrix $\boldsymbol{X}$, with the elements of $\boldsymbol{x} $  read sequentially and written into the matrix column-wise; 2) Perform an N-point IDFnT transform on each row of the signal matrix $\boldsymbol{X}$ to obtain the signal matrix $\boldsymbol{S}$; 3)  Extract elements from matrix  $\boldsymbol{S}$ column-wise and convert them into column vectors $\boldsymbol{s}$. 

 It is worth noting that, the $N$-point IDFnT can be equivalently implemented with low complexity using inverse fast Fourier transform (IFFT), when $N$ is a power of two. Specifically, the transform matrix of the $N$-point IDFnT can be represented as 
\begin{equation} \label{eqn:FiH_LowComplexity}
	\mathbf{\Phi}_N^H=\mathbf{\Theta}_1^H\mathbf{F}_N^H\mathbf{\Theta}_2^H,
\end{equation}
where $\mathbf{F}_N^H$ is the transform matrix of  $N$-point inverse discrete Fourier transform (IDFT), with $\left[ \mathbf{F}_N^H \right] _{m,n}=\frac{1}{\sqrt{N}}e^{j2\pi mn/N}, m,n = 0, \dots, N-1$; $\mathbf{\Theta }_1$ and $\mathbf{\Theta }_2$ are diagonal matrices generated by vectors $\boldsymbol{\theta}_1$ and $\boldsymbol{\theta}_2$, respectively. Namely, $\mathbf{\Theta }_1=\mathrm{diag}\left( \boldsymbol{\theta }_1 \right) 
$ and $\mathbf{\Theta }_2=\mathrm{diag}\left( \boldsymbol{\theta }_2 \right) $, where
the $m$th elements of vectors $\boldsymbol{\theta }_1$ and $\boldsymbol{\theta }_2$ are
\begin{equation}
	\left[ \boldsymbol{\theta}_1 \right] _{m}=\,\,e^{-j\frac{\pi}{4}}e^{j\frac{\pi}{N}m^2},
\end{equation}
and
\begin{equation}
	\left[ \boldsymbol{\theta}_2 \right] _{m}=e^{j\frac{\pi}{N}m^2},
\end{equation}
respectively. Since the IDFT can be implemented using the IFFT, the computational complexity of an $N$-point IDFT is $\mathcal{O} \left( N\log _2N \right)$. Similarly, since the DFnT transform matrix can be expressed as $\mathbf{\Phi }_N=\mathbf{\Theta }_2\mathbf{F}_N\mathbf{\Theta }_1$, this transform can also be efficiently implemented using the FFT.  

\subsection{Channel Model}
\begin{figure}[!ht]
	\includegraphics[width=8cm]{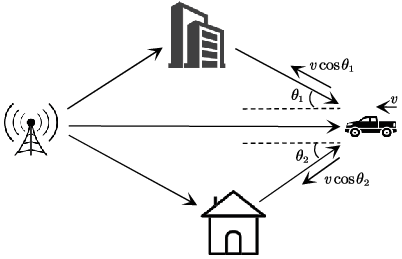}
	\caption{\small The LTV channel with multiple lags and multiple Doppler shifts.}
	\label{fig:mlmd_chnnl}
\end{figure}

In many wireless communication applications, the relative motion between the transmitter and the receiver causes the channel to fluctuate significantly even within the duration of a single signal block. In such cases, the channel should be modeled as a linear time-variant (LTV) system. To better illustrate the mathematical model of this type of LTV channel, we consider the scenario depicted in Fig.  \ref{fig:mlmd_chnnl}, where the receiver moves towards the stationary transmitter at a speed $v$. The propagation environment includes several reflectors. Using the ray-tracing technique, the received signal can be modeled as a superposition of transmitted signal copies from various paths, each characterized by a specific attenuation coefficient, delay, and Doppler shift \cite{LTVChnnlBook}. The Doppler shift experienced by the signal along the $i$th path is 
\begin{equation}
v_i = f_cv\cos \theta _i/C,
\end{equation}
where $C$ represents the propagation speed of the wireless medium (radio waves or acoustic waves); $f_c$ is the carrier frequency of the system;  $\theta _i$ is the angle between the ray of the $i$th path and the direction of relative motion between the transmitter and receiver.
Let's denote the attenuation coefficient and delay of the $i$th path in the equivalent baseband channel as $h_i$ and $\tau_i$, respectively. Then, the equivalent complex baseband signal at the receiver can be represented as \cite{LTVChnnlBook}:
\begin{align} \label{eqn:rttilt}
	r\left( t \right)  = \sum_{i=1}^P{h_i\tilde{s}\left( t-\tau _i \right)}e^{j2\pi v_i\left( t-\tau _i \right)}+\omega\left( t \right) , t\in \left[ 0,T \right] , 
\end{align}
 where $P$ is the number of propagation paths, $\omega\left( t \right)$ is modeled as complex additive white Gaussian noise (AWGN), independent of the transmit signal and the channel, with a mean of zero and a power spectral density of $N_0$.  
 
Considering that the resolution of the sampling interval $Ts$ is usually high enough to approximate the path delay as an integer multiple of the sampling interval in a typical communication system \cite{TseBook}, we can express the delay of the $i$th path as
\begin{equation} \label{eqn:tau}  
\tau _i=l_i T_s,
\end{equation}
where $l_i$ is an integer, and can be seen as the path delay in the  discrete-time  domain.

Let us define the frequency-domain sampling interval of the signal as $\Delta f= 1/T$. We can then express the Doppler frequency shift of the $i$th path as the sum of two components: one that is an integer multiple of $\Delta f$ and another that is a fractional multiple of $\Delta f$, namely,
\begin{equation} \label{eqn:Doppler}
v_i= (k_i+\kappa _i)\Delta f
\end{equation}
where $k_i\in \mathcal{Z}$ and $\kappa _i\in \left(-0.5, 0.5 \right]$. 

Now, by substituting (\ref{eqn:tau}) and  (\ref{eqn:Doppler}) into equation  (\ref{eqn:rttilt}), the received complex baseband signal can be rewritten as 
\begin{equation}
r\left( t \right) =\sum_{i=1}^P{h_i\tilde{s}\left( t-l_iT_s \right) e^{j2\pi \left( k_i+\kappa _i \right) \Delta f\left( t-l_iT_s \right)}}+\omega \left( t \right). 
\end{equation}

Next, let's sample the equivalent baseband signal $r(t)$ at the ideal sampling interval $T_s$. With the CP removed, the corresponding discrete-time signal can be expressed as
\begin{align} \label{eqn:r_n}
r\left[ n \right] =&\sqrt{T_s}\left. r\left( t \right) \right|_{t=nT_s} \nonumber
\\
=&\sqrt{T_s}\sum_{i=1}^P{h_i\tilde{s}\left( nT_s-l_iT_s \right) e^{j2\pi \left( k_i+\kappa _i \right) \Delta f\left( n-l_i \right) T_s}} \nonumber
\\
&+\sqrt{T_s}\omega \left( nT_s \right)  \nonumber
\\
=&\sum_{i=1}^P{\tilde{h}_ie^{j\frac{2\pi}{MN}\left( k_i+\kappa _i \right) n}\tilde{s}\left[ n-l_i \right]}+\omega \left[ n \right] \nonumber
\\
=&\sum_{i=1}^P{\tilde{h}_ie^{j\frac{2\pi}{MN}\left( k_i+\kappa _i \right) n}s\left[ \left< n-l_i \right> _{MN} \right]}+\omega \left[ n \right] , \nonumber
\\
&n=0,1,\dots ,MN-1,
\end{align}
where $\tilde{h}_i$ is the equivalent path gain of the $i$th path,  
\begin{equation}
\tilde{h}_i=h_ie^{-j\frac{2\pi}{MN}\left( k_i+\kappa _i \right) l_i},
\end{equation}
 $\omega\left[n\right]$ represents a discrete-time AWGN sequence, where each sample is independently and identically distributed (IID) with a mean of zero and a variance of $N_0$.
Represent the received discrete-time baseband complex signal and the included AWGN noise as vectors
\begin{equation}
\boldsymbol{r}=\left[ r\left[ 0 \right] ,r\left[ 1 \right] ,\dots ,r\left[ MN-1 \right] \right] ^T,
\end{equation}
and 
\begin{equation}
\boldsymbol{\omega}=\left[ \omega \left[ 0 \right] ,\omega \left[ 1 \right] ,\dots ,\omega \left[ MN-1 \right] \right] ^T,
\end{equation}respectively. Then, the time-domain input-output relationship of the channel described in (\ref{eqn:r_n}) can be rewritten as 
\begin{equation} \label{eqn:v_rcvsig}
\boldsymbol{r}=\mathbf{H}\boldsymbol{s} + \boldsymbol{\omega},
\end{equation}
where 
\begin{equation} \label{eqn:v_chnnl}
\mathbf{H}=\sum_{i=1}^P{\tilde{h}_i\mathbf{\Lambda }^{k_i+\kappa _i}}\mathbf{\Pi }^{l_i}
\end{equation}
is the channel matrix in the time domain, with $\mathbf{\Lambda }$ being a constant $MN\times MN$ diagonal matrix defined by 
\begin{equation} \label{eqn:def_Delta}
\mathbf{\Lambda }=\mathrm{diag}\left( \left[ e^{j\frac{2\pi}{MN}\cdot 0},e^{j\frac{2\pi}{MN}\cdot 1},\dots ,e^{j\frac{2\pi}{MN}\cdot \left( MN-1 \right)} \right] \right) ,
\end{equation}
and $\mathbf{\Pi }$ being an $MN\times MN$ permutation matrix expressed by
\begin{equation}  \label{eqn:def_PI}
\mathbf{\Pi }=\left[ \begin{matrix}
0&		\cdots&		0&		1\\
1&		\ddots&		0&		0\\
\vdots&		\ddots&		\ddots&		\vdots\\
0&		\cdots&		1&		0\\
\end{matrix} \right]. 
\end{equation}
The function of matrix $\mathbf{\Pi }$ is that, when $\mathbf{\Pi }$ multiplies a column vector $\boldsymbol{s}$ of length $MN$, the resulting vector  is a version of $\boldsymbol{s}$ cyclically shifted forward by 1 element, i.e.,  $\mathbf{\Pi }\boldsymbol{s}=\left[ \left[ \boldsymbol{s} \right] _{MN-1},\left[ \boldsymbol{s} \right] _0,\cdots ,\left[ \boldsymbol{s} \right] _{MN-2} \right] ^T.$  On the other hand, when a signal vector is left-multiplied by the diagonal matrix $\mathbf{\Lambda }^{k_i+\kappa _i}$, it undergoes a Doppler shift, with a normalized value $k_i+\kappa _i$. Therefore, using Equations (\ref{eqn:v_rcvsig}) and (\ref{eqn:v_chnnl}), we can consider the received signal vector $\boldsymbol{r}$ as the superposition of $P$ versions of the transmitted signal vector $\boldsymbol{s}$, each version corresponding to an independent path. For the $i$th path, the transmitted signal vector $\boldsymbol{s}$ is first cyclically shifted forward by $l_i$ elements, then experience a Doppler shift with a normalized value $k_i+\kappa _i$, and finally multiplied by an attenuation coefficient $\tilde{h}_i$.

\section{Input-output Relation}
In this section, we first transform the received time-domain signal vector into the GF-domain, obtaining the vector $\boldsymbol{y}$. Then, we derive the input-output relationship of the system in the GF domain, specifically the relationship between the transmitted signal vector $\boldsymbol{x}$ and the received signal vector $\boldsymbol{y}$, and obtain the GF-domain channel matrix. We observe that the GF-domain channel exhibits significant sparsity. By leveraging this sparsity, we propose a low-complexity method for calculating the GF-domain channel matrix.
\subsection{General Expression of the GF-domain Channel Matrix}
Based on (\ref{eqn:v_rcvsig}), we perform a GDFnT on the received time-domain signal vector $\boldsymbol{r}$, obtaining the GF-domain received signal $\boldsymbol{y}$ as
\begin{align}
	\boldsymbol{y}&=\mathbf{\Theta }_{M,N}\boldsymbol{r} \nonumber \\
	&=\mathbf{H}_{\mathrm{eff}}\boldsymbol{x}+\tilde{\boldsymbol{\omega}}, 
\label{eqn:sysmodel-1}
\end{align}
with
\begin{align} 
\mathbf{H}_{\mathrm{eff}}=\mathbf{\Theta }_{M,N}\mathbf{H\Theta }_{M,N}^{H}, \label{eqn:Heff_ver1}
\end{align}
and
\begin{align}
\tilde{\boldsymbol{\omega}}=\mathbf{\Theta }_{M,N}\boldsymbol{\omega },
\end{align}
where $\mathbf{H}_{\mathrm{eff}} $ and $\tilde{\boldsymbol{\omega}}$
 is the equivalent channel matrix and the noise vector in the GF-domain, respectively.  Since matrix $\mathbf{\Theta }_{M,N}$ is unitary and the elements of noise vector $\boldsymbol{\omega }$ are modeled as IID AWGN, the elements of $\tilde{\boldsymbol{\omega}}$ are also IID AWGN with zero mean and variance $N_0$.
 
Substituting (\ref{eqn:v_chnnl}) into (\ref{eqn:Heff_ver1}), we have
\begin{align}
\mathbf{H}_{\mathrm{eff}}\,\,&=\,\mathbf{\Theta }_{M,N}\mathbf{H\Theta }_{M,N}^{H}
\\
&=\sum_{i=1}^P{\tilde{h}_i\,\mathbf{\Theta }_{M,N}\mathbf{\Lambda }^{k_i+\kappa _i}\mathbf{\Pi }^{l_i}\mathbf{\Theta }_{M,N}^{H}}
\\
&=\sum_{i=1}^P{\tilde{h}_i\mathbf{P}^{\left( i \right)}\mathbf{Q}^{\left( i \right)}}
 \label{H_eff_2}
\end{align}
with
\begin{equation} \label{eqn:P_i}
\mathbf{P}^{\left( i \right)}=\mathbf{\Theta }_{M,N}\mathbf{\Lambda }^{k_i+\kappa _i}\mathbf{\Theta }_{M,N}^{H}
\end{equation}
and
\begin{equation} \label{eqn:Q_i}
\mathbf{Q}^{\left( i \right)}=\mathbf{\Theta }_{M,N}\mathbf{\Pi }^{l_i}\mathbf{\Theta }_{M,N}^{H}.
\end{equation}
\subsection{Complexity-reduced Computation of the GF-domain Channel Matrix}
 Note that directly computing the channel matrix using equation (\ref{H_eff_2}) results in extremely high computational complexity. This high complexity arises from the multiple multiplications of $MN \times MN$ dimensional matrices required in the calculation process.To obtain a computation-efficient and intuitive expression for the GF-domain channel matrix $\mathbf{H}_{\mathrm{eff}}$, we simplify the expressions of matrices $\mathbf{P}^{\left( i \right)}$ and $\mathbf{Q}^{\left( i \right)}$, respectively.

To simplify (\ref{eqn:Q_i}), we need the following lemma.
\begin{lemma} \label{Lemma_1}
	Matrices $\mathbf{\Phi }_{M,N}^{\mathcal{H}}$ and $\mathbf{\Pi }$ satisfy the commutative property of multiplication, i.e., $\mathbf{\Pi \Theta }_{M,N}^{H}\,\,=\,\,\mathbf{\Theta }_{M,N}^{H}\mathbf{\Pi }.$
\end{lemma}

\emph{Proof}: See Appendix A.

By applying Lemma \ref{Lemma_1} and using mathematical induction, it can be readily proven that equation (\ref{eqn:Q_i}) simplifies to
\begin{equation}
	\mathbf{Q}^{\left( i \right)}=\mathbf{\Pi }^{l_i}. \label{Q_i_item}
\end{equation}

Next, we derive the expression for matrix   $\mathbf{P}^{\left( i \right)}$ in two steps. First, we derive the expression for the case when the normalized Doppler shift is an integer. Subsequently, we extend this result to scenarios where the normalized Doppler shift includes a fractional component. 

\emph{1) Case of integer Doppler shifts} 

Using (\ref{eqn:def_Delta}), we can express the diagonal matrix $\mathbf{\Lambda }^{k_i}$ in the form of a block matrix as
\begin{equation} \label{mat:Lambda}
	\mathbf{\Lambda }^{k_i}=\left[ \begin{matrix}
		\mathbf{\Lambda }_{0}^{k_i}&		&		&		\\
		&		\mathbf{\Lambda }_{1}^{k_i}&		&		\\
		&		&		\ddots&		\\
		&		&		&		\mathbf{\Lambda }_{N-1}^{k_i}\\
	\end{matrix} \right] 
\end{equation}
with
\begin{align} \label{Lambda_subblock}
\mathbf{\Lambda}_{n}^{k_i}&=\mathrm{diag}\left( \left[ e^{j\frac{2\pi}{MN}k_i\left( nM+0 \right)},e^{j\frac{2\pi}{MN}k_i\left( nM+1 \right)},\cdots , \right. \right. \nonumber \\
& \left. \left. \qquad \qquad e^{j\frac{2\pi}{MN}k_i\left( nM+M-1 \right)} \right] \right) 
\nonumber \\
&=e^{j\frac{2\pi}{N}k_in}\mathring{\mathbf{\Lambda}}^{k_i}, 
\end{align}
\begin{equation}
\mathring{\mathbf{\Lambda}}\triangleq \mathrm{diag}\left( \left[ e^{j\frac{2\pi}{MN}0},e^{j\frac{2\pi}{MN}1},\cdots ,e^{j\frac{2\pi}{MN}\left( M-1 \right)} \right] \right), \label{Lambda_basic}
\end{equation}
and $n=0,1,\dots,N-1$.

According to (\ref{Phi_n_n}), (\ref{blockMat_theta}), (\ref{eqn:P_i}), (\ref{mat:Lambda}) and (\ref{Lambda_subblock}), the matrix $\mathbf{P}^{\left( i \right)}$ can also be expressed as a block matrix in the form 
\begin{align}
	\mathbf{P}^{\left( i \right)}=\left[ \begin{matrix}
		\mathbf{P}_{0,0}^{\left( i \right)}&		\mathbf{P}_{0,1}^{\left( i \right)}&		\cdots&		\mathbf{P}_{0,N-1}^{\left( i \right)}\\
		\mathbf{P}_{1,0}^{\left( i \right)}&		\mathbf{P}_{1,1}^{\left( i \right)}&		\cdots&		\mathbf{P}_{1,N-1}^{\left( i \right)}\\
		\vdots&		\vdots&		\ddots&		\vdots\\
		\mathbf{P}_{N-1,0}^{\left( i \right)}&		\mathbf{P}_{N-1,1}^{\left( i \right)}&		\cdots&		\mathbf{P}_{N-1,N-1}^{\left( i \right)}\\
	\end{matrix} \right] 
\end{align}
with 
\begin{align}
	\mathbf{P}_{n,n^{\prime}}^{\left( i \right)}&=\sum_{\bar{n}=0}^{N-1}{\left[ \mathbf{\Phi }_N \right] _{n,\bar{n}}\left[ \mathbf{\Phi }_{N}^{H} \right] _{\bar{n},n^{\prime}}e^{j\frac{2\pi}{N}\bar{n}}\mathring{\mathbf{\Lambda}}^{k_i}}
	\nonumber \\
	&=\frac{1}{N}e^{j\frac{\pi}{N}\left[ n^2-{n^{\prime}}^2 \right]}\left( \sum_{\bar{n}=0}^{N-1}{e^{j\frac{2\pi}{N}\left( n^{\prime}-n+k_i \right) \bar{n}}} \right) \mathring{\mathbf{\Lambda}}^{k_i}
	\nonumber \\
	&=e^{j\frac{\pi}{N}\left[ n^2-{n^{\prime}}^2 \right]}\delta \left[ \left< n-n^{\prime}-k_i \right> _N \right] \mathring{\mathbf{\Lambda}}^{k_i} \label{P_subBlock}
\end{align}
and $n,n^\prime = 0,1,\dots,N-1$. 

Next, we derive the expression for the $(p,p^\prime)$th element of matrix
\begin{align}
\left[ \mathbf{P}^{\left( i \right)} \right] _{p,p^{\prime}}&=e^{j\frac{\pi}{N}\left[ n^2-{n^{\prime}}^2 \right]}\delta \left[ \left< n-n^{\prime}-k_i \right> _N \right] \left[ \mathring{\mathbf{\Lambda}}^{k_i} \right] _{m,m^{\prime}} \nonumber
\\
&=e^{j\frac{\pi}{N}\left[ n^2-{n^{\prime}}^2 \right]}e^{j\frac{2\pi}{MN}k_im}\delta \left[ \left< n-n^{\prime}-k_i \right> _N \right] \nonumber \\
   &\quad \cdot \delta \left[ m-m^{\prime} \right]  \nonumber
\\
&=e^{j\frac{\pi}{N}\left[ \lfloor \frac{p}{M} \rfloor ^2-\lfloor \frac{p^{\prime}}{M} \rfloor ^2 \right]}e^{j\frac{2\pi}{MN}\left( k_i+b \right) \left< p \right> _M}  \nonumber
\\
& \quad \cdot \delta \left[ \left< p-p^{\prime}-k_iM \right> _{MN} \right].  \label{P_withIntegerShift}
\end{align}
\emph{2) Case of Fractional Doppler shifts}

Given a non-zero fractional Doppler shift $\kappa _i$, we can write the diagonal matrix $
\mathbf{\Lambda }^{\kappa _i}$ as 
\begin{equation}
	\mathbf{\Lambda }^{\kappa _i}=\mathrm{diag}\left( \boldsymbol{v}_i \right),
\end{equation}
with 
\begin{equation}
	\boldsymbol{v}_i=\left[ e^{j\frac{2\pi}{MN}\kappa _i\cdot 0},e^{j\frac{2\pi}{MN}\kappa _i\cdot 1},\dots ,e^{j\frac{2\pi}{MN}\kappa _i\cdot \left( MN-1 \right)} \right] ^T.
\end{equation}

Let us define a set of complex vectors 
\begin{equation}
\mathcal{V} =\left\{ \boldsymbol{\vartheta }_b|b=-\frac{MN}{2},\dots ,0,\dots ,\frac{MN}{2}-1 \right\},
\end{equation}
where
\begin{equation}
	\boldsymbol{\vartheta }_b\,\,=\,\,\left[ e^{j\frac{2\pi}{MN}b\cdot 0},e^{j\frac{2\pi}{MN}b\cdot 1},\dots ,e^{j\frac{2\pi}{MN}b\cdot \left( MN-1 \right)} \right] ^T.
\end{equation}
It should be noted that the set $\mathcal{V}$ constitutes an orthonormal basis in the $MN$-dimensional vector space over the field of complex numbers. Then we can express the vector $\boldsymbol{v}_i$ as a linear combination of the vectors in the orthonormal basis $\mathcal{V}$, i.e.,
\begin{equation}
	\boldsymbol{v}_i=\sum_{b=-MN/2}^{MN/2-1}{\lambda _{i,b}\boldsymbol{\vartheta }_b}
\end{equation}
where
\begin{align}
\lambda _{i,b}&=\frac{\boldsymbol{\vartheta }_{b}^{H}\boldsymbol{v}_i}{\boldsymbol{\vartheta }_{b}^{H}\boldsymbol{\vartheta }_b} \nonumber
	\\
	&=\frac{1}{MN}\sum_{n=0}^{MN-1}{e^{j\frac{2\pi}{MN}\kappa _in}}e^{-j\frac{2\pi}{MN}bn} \nonumber
	\\
	&=\frac{1}{MN}\frac{e^{j2\pi \kappa _i}-1}{e^{j\frac{2\pi}{MN}\left( \kappa _i-b \right)}-1}. \label{lambda_i_b}
\end{align}

In equation (\ref{lambda_i_b}), as the absolute value of $b$ increases from zero,  the magnitude of the denominator $
\left| e^{j\frac{2\pi}{MN}\left( \kappa _i-b \right)}-1 \right|$ in the fraction also increases from a value close to zero. Consequently, the magnitude of $\lambda _{i,b}$ decreases rapidly with the increasing absolute value of 
$b$. Therefore, $\boldsymbol{\varLambda }^{\kappa _i}$ can be approximated as
\begin{align}
	\boldsymbol{\varLambda }^{\kappa _i} &\approx \sum_{b=-B_i}^{B_i}{\lambda _{i,b}\mathrm{diag}\left( \boldsymbol{v}_b \right)} \nonumber
	\\
	&\approx \sum_{b=-B_i}^{B_i}{\lambda _{i,b}\boldsymbol{\varLambda }^b}. \label{fractional_eqv}
\end{align}
whee $B_i$ is a constant integer that controls the approximation accuracy.

Based on Equation (\ref{fractional_eqv}), it can be observed that a fractional Doppler shift can be approximately equivalent to multiple integer Doppler shifts. Therefore, using the result given in (\ref{P_withIntegerShift}), we can approximate $\mathbf{P}^{\left( i \right)}$ with path Doppler shift $
k_i+\kappa _i$ as
\begin{align}
	\left[ \tilde{\mathbf{P}}^{\left( i \right)} \right] _{p,p^{\prime}}=&\sum_{b=-B_i}^{B_i}{\lambda _{i,b}e^{j\frac{\pi}{N}\left[ n^2-{n^{\prime}}^2 \right]}e^{j\frac{2\pi}{MN}\left( k_i+b \right) m}} 
	\nonumber \\
	& \qquad \quad \cdot \delta \left[ \left< n-n^{\prime}-k_i-b \right> _N \right] \delta \left[ m-m^{\prime} \right] \nonumber \\
	=&\sum_{b=-B_i}^{B_i}{\lambda _{i,b}e^{j\frac{\pi}{N}\left[ \lfloor \frac{p}{M} \rfloor ^2-\lfloor \frac{p^{\prime}}{M} \rfloor ^2 \right]}e^{j\frac{2\pi}{MN}\left( k_i+b \right) \left< p \right> _M}} \nonumber\\
	& \qquad \cdot \delta \left[ \left< p-p^{\prime}-\left( k_i+b \right) M \right> _{MN} \right].
	\label{P_i_item}
\end{align}

According to (\ref{H_eff_2}), (\ref{Q_i_item}), and (\ref{P_i_item}), the generalized-Fresnel domain channel matrix $\mathbf{H}_{\mathrm{eff}}$ can be approximated as $\tilde{\mathbf{H}}_{\mathrm{eff}}$, whose ($p,p\prime$)th element is 
\begin{align} \label{H_tild_p_p}
	\left[ \tilde{\mathbf{H}}_{\mathrm{eff}} \right] _{p,p^{\prime}}&=\sum_{i=1}^P{\tilde{h}_i\left[ \tilde{\mathbf{P}}^{\left( i \right)}\mathbf{Q}^{\left( i \right)} \right] _{p,p^{\prime}}} \nonumber
	\\
	&=\sum_{i=1}^P{\tilde{h}_i\left[ \tilde{\mathbf{P}}^{\left( i \right)} \right] _{p,\left< p^{\prime}+l_i \right> _{MN}}} \nonumber
	\\
	&=\,\,\sum_{i=1}^P{\sum_{b=-B_i}^{B_i}{\left(\tilde{h}_i\lambda _{i,b}e^{j\frac{\pi}{N}\left[ \lfloor \frac{p}{M} \rfloor ^2-\lfloor \frac{p^{\prime}+l_i}{M} \rfloor ^2 \right]}\right.}} \nonumber \\
	&\left.e^{j\frac{2\pi}{MN}\left( k_i+b \right) \left< p \right> _M}\delta \left[ \left< p-p^{\prime}-l_i-\left( k_i+b \right) M \right> _{MN} \right]\right). \nonumber \\
	& \qquad  
\end{align}
Note that, in (\ref{H_tild_p_p}), $B_i$ and $\lambda _{i,0}$ should be set  to zero and 1, respectively, when $\kappa_i = 0$.

Based on (\ref{H_tild_p_p}), we can consider the GF-domain channel as a channel comprising $\sum_{i=1}^P{\left( 2B_i+1 \right)}$ \emph{virtual paths}, where the $(i,b)$th virtual path induces a cyclic shift of $l_i-\left( k_i+b \right) M$ positions on each transmitted symbol (element of $\boldsymbol{x}$). Let us denote the set of all virtual path indices as set
\begin{equation}
	\mathcal{A} =\left\{ \left( i,b \right) |i=0,\cdots ,P-1, b=-B_i,\cdots ,0,\cdots ,B_i \right\}.
\end{equation}
It is evident that different virtual paths may have the same number of cyclic shifts. We group all virtual paths with the same number of cyclic shifts together, denoting their indices as set 
\begin{align}
	\mathcal{A} _{\ell}=\left\{ \left( i,b \right) |l_i-\left( k_i+b \right) M=d_{\ell}, \left( i,b \right) \in \mathcal{A} \right\}  \nonumber \\
	\ell = 0,1,\cdots,L-1, \label{A_L}
\end{align}
where $L$ is the number of groups, $d_\ell$ represents the number of positions for the cyclic shift of the virtual paths corresponding to $\mathcal{A_\ell}$. All indices sets expressed in (\ref{A_L}) are mutually disjoint, and their union forms the set $\mathcal{A}$, i.e.,  $
d_{\ell _1}\ne d_{\ell _2}\,\,,\ell _1\ne \ell _2$ and $\mathcal{A} =\bigcup_{\ell =0}^{L-1}{\mathcal{A} _{\ell}}$.

According to (\ref{H_tild_p_p}) and (\ref{A_L}), the $(p,p^\prime)$th element of $\tilde{\mathbf{H}}_{\mathrm{eff}}$ can be rewritten as
\begin{align} \label{H_p_p_sparse}
\left[ \tilde{\mathbf{H}}_{\mathrm{eff}} \right] _{p,p^{\prime}}=&\sum_{\ell =1}^L{\breve{h}_{p,p^{\prime}}^{\ell}\delta \left[ \left< p-p^{\prime}-d_{\ell} \right> _{MN} \right]} \nonumber
	\\
	=&\begin{cases}
		\breve{h}_{p,p^{\prime}}^{\ell},&p=\left< p^{\prime}+d_{\ell} \right> _M, \ell =0,\cdots ,L-1\\
		0,&		\mathrm{others}\\
	\end{cases}
\end{align}
where
\begin{equation}
	\breve{h}_{p,p^{\prime}}^{\ell}=\sum_{\left( i,b \right) \in \mathcal{A} _{\ell}}{\tilde{h}_i\lambda _{i,b}e^{j\frac{\pi}{N}\left[ \lfloor \frac{p}{M} \rfloor ^2-\lfloor \frac{p^{\prime}+l_i}{M} \rfloor ^2 \right]}e^{j\frac{2\pi}{MN}\left( k_i+b \right) \left< p \right> _M}}.
\end{equation}

From (\ref{H_p_p_sparse}), it can be seen that $\tilde{\mathbf{H}}_{\mathrm{eff}}$ is a sparse matrix, with each row (or column) containing $L$ nonzero elements. Therefore, in practical applications, the GF-domain channel matrix can be obtained by directly calculating the nonzero elements of $\tilde{\mathbf{H}}_{\mathrm{eff}}$, significantly reducing computational overhead compared to calculating it with  (\ref{H_eff_2}). 

\section{Message Passing Based Detector} \label{sec:Detector}
In this section, we assume that the path parameters of the equivalent baseband channel, including time delays, Doppler shifts, and attenuation coefficients, are perfectly known at the receiver. We propose an iterative detector using the MP algorithm, which leverages the sparsity of the GF-domain channel matrix $\tilde{\mathbf{H}}_{\mathrm{eff}}$.

Considering the noise introduced by the approximate computation of the GF-domain channel matrix, based on (\ref{eqn:sysmodel-1}), we rewrite the GF-domain input-output relationship as
\begin{equation}
	\boldsymbol{y}=\tilde{\mathbf{H}}_{\mathrm{eff}}\boldsymbol{x}+\breve{\boldsymbol{\omega}},
\end{equation}
where $\breve{\boldsymbol{\omega}}$ represents the noise vector encompassing both channel additive noise and channel approximation noise. For simplicity, we assume that the elements of $\breve{\boldsymbol{\omega}}$ are i.i.d. AWGN with a mean of zero and a variance of $\sigma_{0}^2$.

From (\ref{H_p_p_sparse}), it is evident that the received symbol $\left[ \boldsymbol{y} \right] _p$, $p=0,1,\cdots, MN-1$, contains contributions from $L$ transmitted symbols whose indices can be expressed by vector
\begin{equation}
\boldsymbol{b}_p=\left< \left[ p-d_0, p-d_1,\cdots ,p-d_{L-1} \right] \right> _{MN}.
\end{equation}
On the other hand, the  transmitted symbol $\left[ \boldsymbol{x} \right] _{p\prime}$, $
p^{\prime}=0,1,\cdots ,MN-1$, affects $L$ received symbols whose indices are expressed by vector
\begin{equation}
	\boldsymbol{q}_{p^{\prime}}=\left< \left[ p^{\prime}+d_0, p^{\prime}+d_1,\cdots ,p^{\prime}+d_{L-1} \right] \right> _{MN}.
\end{equation}

Next, we consider the received symbol $\left[ \boldsymbol{y} \right] _p$ as an observation of the transmitted symbol $\left[ \boldsymbol{x} \right] _{\left[ \boldsymbol{b}_p \right] _{\ell}}$ and express their relationship as
\begin{align} \label{ISI_model}
\left[ \boldsymbol{y} \right] _p=&{\left[ \tilde{\mathbf{H}}_{\mathrm{eff}} \right] _{p,}}_{\left[ \boldsymbol{b}_p \right] _{\ell}}\cdot \left[ \boldsymbol{x} \right] _{\left[ \boldsymbol{b}_p \right] _{\ell}} \nonumber
\\
&+\mathop {\underbrace{\sum_{\begin{array}{c}
				i=0\\
				i\ne \ell\\
		\end{array}}^{L-1}{{\left[ \tilde{\mathbf{H}}_{\mathrm{eff}} \right] _{p,}}_{\left[ \boldsymbol{b}_p \right] _i}\cdot \left[ \boldsymbol{x} \right] _{\left[ \boldsymbol{b}_p \right] _i}+\left[ \breve{\boldsymbol{\omega}} \right] _p}}} \limits_{\left[ \mathbf{W} \right] _{p,\left[ \boldsymbol{b}_p \right] _{\ell}}},
\end{align}
where $\left[ \mathbf{W} \right] _{p,\left[ \boldsymbol{b}_p \right] _{\ell}}$ represents the observation noise, which includes Gaussian white noise $\left[ \breve{\boldsymbol{\omega}} \right] _p$ and interference caused by the other $L-1$ transmitted symbols. For computational simplicity, we assume that the elements of $\boldsymbol{x}$ are independent with each other, and $\left[ \mathbf{W} \right] _{p,\left[ \boldsymbol{b}_p \right] _{\ell}}$ follows a  complex Gaussian distribution. 

From (\ref{ISI_model}), it is evident that if the discrete probability distribution of the random vector $\boldsymbol{x}$ is known, the mean and variance of the observation noise $\left[ \mathbf{W} \right] _{p,\left[ \boldsymbol{b}_p \right] _{\ell}}$, $p=0,\cdots,MN-1$, $\ell=0,\cdots,L-1$, can be obtained. Conversely, if the mean and variance of $\left[ \mathbf{W} \right] _{p,\left[ \boldsymbol{b}_p \right] _{\ell}}$ is known, the probability distribution of $\left[ \boldsymbol{x} \right] _{\left[ \boldsymbol{b}_p \right] _{\ell}}$ can be calculated from the observed value $\left[ \boldsymbol{y} \right] _p$. Therefore, the estimation problem for $\boldsymbol{x}$ and $\left[ \mathbf{W} \right] _{p,\left[ \boldsymbol{b}_p \right] _{\ell}}$ represents a classic ``chicken-and-egg'' problem, which can be resolved using an iterative estimation method based on message passing \cite{OTFS-HongYi}.

\begin{figure}[!htbp]
	\centering
	\includegraphics[width=8cm]{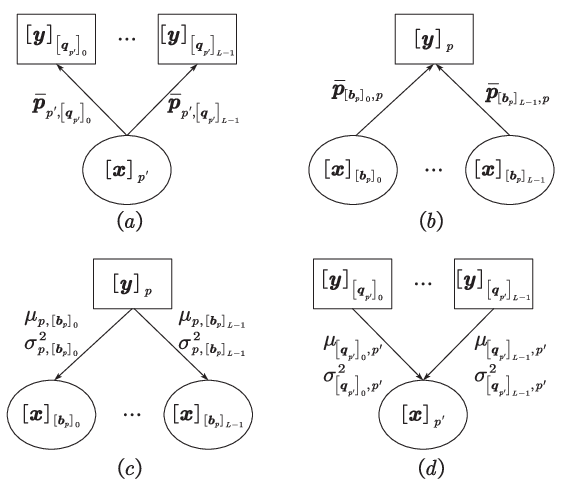}
	\caption{\small Messages in  the  factor graph: (a) messages sent from a variable node to $L$ observation nodes, (b)  messages got by  an observation node from $L$ variable nodes, (c) messages sent from an observation node to $L$ variable nodes, (d) messages got by a variable node from $L$ observation nodes.}
	\label{fig:factorGraph}
\end{figure}

First, we depict the factor graph in Fig. (\ref{fig:factorGraph}). In the factor graph, the variable nodes, representing the transmitted symbols are illustrated as ellipses, while the factor nodes, representing the observed signal, are illustrated as rectangles. The connections between variable nodes and factor nodes are sparse. As shown in subfigure (a), the variable node $\left[ \boldsymbol{x} \right] _{p^{\prime}}$ sends the probability distribution of $\left[ \boldsymbol{x} \right] _{p^{\prime}}$, i.e., $\bar{\boldsymbol{p}}_{p^{\prime},\left[ \boldsymbol{q}_{p^{\prime}} \right] _{\ell}}$, to the factor node $\left[ \boldsymbol{y} \right] _{\left[ \boldsymbol{q}_{p^{\prime}} \right] _{\ell}}$. $\bar{\boldsymbol{p}}_{p^{\prime},\left[ \boldsymbol{q}_{p^{\prime}} \right] _{\ell}}$ is a probability mass function (pmf) defined as
\begin{align}
\bar{\boldsymbol{p}}_{p^{\prime},\left[ \boldsymbol{q}_{p^{\prime}} \right] _{\ell}}=&\left[ \bar{p}_{p^{\prime},\left[ \boldsymbol{q}_{p^{\prime}} \right] _{\ell}}\left( \alpha _0 \right) ,\bar{p}_{p^{\prime},\left[ \boldsymbol{q}_{p^{\prime}} \right] _{\ell}}\left( \alpha _1 \right) , \right. \nonumber \\
&\left. \cdots ,\bar{p}_{p^{\prime},\left[ \boldsymbol{q}_{p^{\prime}} \right] _{\ell}}\left( \alpha _{\mathcal{M} -1} \right) \right], 
\end{align}
where $\bar{p}_{p^{\prime},\left[ \boldsymbol{q}_{p^{\prime}} \right] _{\ell}}\left( \alpha _m \right)$ is a posterior probability of event $\left[ \boldsymbol{x} \right] _{p^{\prime}}=\alpha _m$, $m=0,1,\cdots,\mathcal{M}-1$. 
As illustrated in subfigure (b), each factor node receives the posterior pmf of the $L$ connected variable nodes from $L$ links. From subfigure (c), we know that a single observation node transmits the mean and variance of the observation noise for each of the $L$ connected variable nodes. Subfigure (d) indicates that each variable node receives the mean and variance of the observation noise from each of the $L$ observation nodes.

\begin{algorithm}[!t]
	\caption{MP Algorithm for GOCDM Symbol Detection}
	\label{alg1}
	\begin{algorithmic}[1]
		\REQUIRE 
		$\boldsymbol{y}$ (the received GF-domain vector), \\
		$\tilde{\mathbf{H}}_{\mathrm{eff}}$ (the GF-domain channel matrix),\\
		$\boldsymbol{b}_p$ and $\boldsymbol{q}_{p^{\prime}}$, $p,p^\prime =0,1,\dots,MN-1$ (indices vectors), \\
		$\sigma _{o}^{2}$ (variance of elements of i.i.d. AWGN $\breve{\boldsymbol{\omega}}$). \\
		$I_{\max}$ (the maximum allowed number of iterations).
		
		\ENSURE  
		$\hat{\boldsymbol{x}}$ (estimation of the transmit symbols) \\
		
		\STATE \textbf{Initiation:} \\
		$\bar{p}_{p^{\prime},\left[ \boldsymbol{q}_{p^{\prime}} \right] _{\ell}}\left( \alpha _m \right) \gets 1/\mathcal{M}$, with $p^\prime=0,\dots,MN-1$, \\
		$\ell=0, \dots, L-1$, and $m=0,\dots,\mathcal{M}-1$. \\ 
		$i \leftarrow 1$ (iteration count), \\
		$\eta \leftarrow 0$ (convergence indicator), \\
		$\eta _{\max} \leftarrow \eta$ (maximum $\eta$ in previous iterations).\\
		
		\WHILE{$\left( i\leqslant I_{\max} \right) \mathrm{and} \left( \eta <1 \right)$}
		\STATE Calculate $\mu _{p,\left[ \boldsymbol{b}_p \right] _{\ell}}$ and $\sigma _{p,\left[ \boldsymbol{b}_p \right] _{\ell}}^{2}$ by (\ref{eqn:mean_awgn}) and (\ref{eqn:var_awgn}), respectively, for  $p=0,\dots,MN-1$,
		$\ell=0, \dots, L-1$.
		\STATE Calculate $\bar{p}_{p^\prime,\left[ \boldsymbol{q}_{p^\prime} \right] _{\ell}}\left( \alpha _m \right) \leftarrow \varDelta \tilde{p}_{p^\prime,\left[ \boldsymbol{q}_{p^\prime} \right] _{\ell}}\left( \alpha _m \right)$ \\
		$+\left( 1-\varDelta \right) \bar{p}_{p^\prime,\left[ \boldsymbol{q}_{p^\prime} \right] _{\ell}}\left( \alpha _m \right)$, with $
		\tilde{p}_{p^\prime,\left[ \boldsymbol{q}_{p^\prime} \right] _{\ell}}\left( \alpha _m \right)$ calculated by (\ref{eqn:tilt_p}), (\ref{eqn:breve_p}), and (\ref{eqn:awgn_pdf}), for  $p^\prime=0,\dots,MN-1$,
		$\ell=0, \dots, L-1$, $m=0,\dots,\mathcal{M}-1$.
		
		\STATE  Calculate convergence indicator $\eta$ by (\ref{eqn:convergence_factor}), (\ref{eqn:combinedProb_xn}), and (\ref{eqn:raw_combinedProb_xn}).
		\IF{$\eta >\eta _{\max}$}
		\STATE $\eta _{\max}\gets \eta$ 
		\STATE Update $\hat{\boldsymbol{x}}$ by (\ref{eqn:x_estimation}).
		\ELSIF{$\eta <\eta _{\max}-\epsilon$} 
		\STATE \textbf{break}
		\ENDIF
		\STATE $i\leftarrow i+1$
		\ENDWHILE
		\RETURN $\hat{\boldsymbol{x}}$
		
	\end{algorithmic}
\end{algorithm}

Based on Figure \ref{fig:factorGraph}, we describe the MP algorithm for variable node estimation in \textbf{Algorithm} \ref{alg1}. We initialize $\bar{p}_{p^\prime,\left[ \boldsymbol{q}_{p^\prime} \right] _{\ell}}\left( \alpha _m \right) $ as $1/\mathcal{M}$ for $p^\prime=0,\dots,MN-1$, $\ell=0, \dots, L-1$, and $m=0,\dots,\mathcal{M}-1$. 
In each iteration, we first calculate the mean and variance of the observation noise at the observation node side, as given by
\begin{equation}
\mu _{p,\left[ \boldsymbol{b}_p \right] _{\ell}}=\sum_{\begin{array}{c}
		i=0\\
		i\ne \ell\\
\end{array}}^{L-1}{\sum_{m=0}^{M-1}{\bar{p}_{\left[ \boldsymbol{b}_p \right] _i,p}\left( \alpha _m \right) {\left[ \tilde{\mathbf{H}}_{\mathrm{eff}} \right] _{p,}}_{\left[ \boldsymbol{b}_p \right] _i}\alpha _m}}, \label{eqn:mean_awgn}
\end{equation}
and
\begin{align}
&\sigma _{p,\left[ \boldsymbol{b}_p \right] _{\ell}}^{2}=\sum_{\begin{array}{c}
		i=0\\
		i\ne \ell\\
\end{array}}^{L-1}{\left( \sum_{m=0}^{M-1}{p_{\left[ \boldsymbol{b}_p \right] _i,p}\left( \alpha _m \right) \left| {\left[ \tilde{\mathbf{H}}_{\mathrm{eff}} \right] _{p,}}_{\left[ \boldsymbol{b}_p \right] _i} \right|^2\left| \alpha _m \right|^2} \right.}
  \nonumber\\
	& \left. -\left| \sum_{m=0}^{M-1}{p_{\left[ \boldsymbol{b}_p \right] _i,p}\left( \alpha _m \right) \alpha _m{\left[ \tilde{\mathbf{H}}_{\mathrm{eff}} \right] _{p,}}_{\left[ \boldsymbol{b}_p \right] _i}} \right|^2 \right) +\sigma _{o}^{2}, \label{eqn:var_awgn}
\end{align}
respectively, for $p=0,\dots,MN-1$, $\ell=0, \dots, L-1$. Then, at the variable node side, the posterior probability of the transmitted symbol $\left[ \boldsymbol{x} \right] _{p^\prime}$ for observation node $\left[ \boldsymbol{y} \right] _{\left[ \boldsymbol{q}_{p^{\prime}} \right] _{\ell}}
$, i.e., $\bar{p}_{p^\prime,\left[ \boldsymbol{q}_{p^\prime} \right] _{\ell}}\left( \alpha _m \right)$, is updated as $\left( 1-\varDelta \right) \bar{p}_{p^{\prime},\left[ \boldsymbol{q}_{p^{\prime}} \right] _{\ell}}\left( \alpha _m \right) +\varDelta \tilde{p}_{p^{\prime},\left[ \boldsymbol{q}_{p^{\prime}} \right] _{\ell}}\left( \alpha _m \right) $, where $\varDelta$ is a constant in the range (0,1). The parameter $\varDelta$, referred to as the \emph{damping factor}, is used to control the convergence rate of the algorithm. The term $\tilde{p}_{p^{\prime},\left[ \boldsymbol{q}_{p^{\prime}} \right] _{\ell}}\left( \alpha _m \right)$ represents the  posterior probability of the transmitted symbol $\left[ \boldsymbol{x} \right] _{p^\prime}$ for observation node $\left[ \boldsymbol{y} \right] _{\left[ \boldsymbol{q}_{p^{\prime}} \right] _{\ell}}$ calculated based on the most recently received messages from the other $L-1$ observation nodes, as given by 
\begin{equation}
\tilde{p}_{p^{\prime},\left[ \boldsymbol{q}_{p^{\prime}} \right] _{\ell}}\left( \alpha _m \right) =\frac{\breve{p}_{p^{\prime},\left[ \boldsymbol{q}_{p^{\prime}} \right] _{\ell}}\left( \alpha _m \right)}{\sum_{k=0}^{M-1}{\breve{p}_{p^{\prime},\left[ \boldsymbol{q}_{p^{\prime}} \right] _{\ell}}\left( \alpha _k \right)}} \label{eqn:tilt_p}
\end{equation}
where
\begin{equation}
\breve{p}_{p^{\prime},\left[ \boldsymbol{q}_{p^{\prime}} \right] _{\ell}}\left( \alpha _m \right) =\prod_{i=0,i\ne \ell}^{L-1}{\bar{p}\left( \left[ \boldsymbol{y} \right] _{_{\left[ \boldsymbol{q}_{p^{\prime}} \right] _i}}|\left[ \boldsymbol{x} \right] _{p^{\prime}}=\alpha _m \right)}, \label{eqn:breve_p}
\end{equation}
with
\begin{align}
&\bar{p}\left( \left[ \boldsymbol{y} \right] _{_{\left[ \boldsymbol{q}_{p^{\prime}} \right] _i}}|\left[ \boldsymbol{x} \right] _{p^{\prime}}=\alpha _m \right) = \nonumber
\\
&\exp \left( \frac{-\left| \left[ \boldsymbol{y} \right] _{\left[ \boldsymbol{q}_{p^{\prime}} \right] _i}-\left[ \tilde{\mathbf{H}}_{\mathrm{eff}} \right] _{\left[ \boldsymbol{q}_{p^{\prime}} \right] _i,p^{\prime}}\left[ \boldsymbol{x} \right] _{p^{\prime}}-\mu _{\left[ \boldsymbol{q}_{p^{\prime}} \right] _i,p^{\prime}} \right|^2}{\sigma _{\left[ \boldsymbol{q}_{p^{\prime}} \right] _i,p^{\prime}}^{2}} \right). \label{eqn:awgn_pdf}
\end{align}
Next, the convergence factor $\eta$ is calculated by
\begin{equation}
\eta =\frac{1}{N}\sum_{n=0}^{N-1}{\mathbb{I} \left( \mathop {\max } \limits_{\alpha _m\in \mathcal{X}}\bar{p}_{p^\prime}\left( \alpha _m \right) \geqslant \gamma \right)}, \label{eqn:convergence_factor}
\end{equation}
where $\gamma$ is a probabilistic threshold constant, which is less than but close to 1; $\mathbb{I} \left( \cdot \right)$ is the indicator function, which evaluates to 1 if the condition inside the parentheses is true and 0 otherwise; $\bar{p}_{p^\prime}\left( \alpha _m \right)$ is the posterior probability of  event $\left[ \boldsymbol{x} \right] _{p^\prime}=\alpha _m$ after obtaining  the messages from all the $L$ related observation nodes. It is calculated by
\begin{equation}
\bar{p}_{p^{\prime}}\left( \alpha _m \right) =\frac{\hat{p}_{p^{\prime}}\left( \alpha _m \right)}{\sum_{k=0}^{M-1}{\hat{p}_{p^{\prime}}\left( \alpha _k \right)}}
\label{eqn:combinedProb_xn}
\end{equation}
with
\begin{equation}
\hat{p}_{p^{\prime}}\left( \alpha _m \right) =\prod_{\ell =0}^{L-1}{\bar{p}\left( \left[ \boldsymbol{y} \right] _{_{\left[ \boldsymbol{q}_{p^{\prime}} \right] _{\ell}}}|\left[ \boldsymbol{x} \right] _{p^{\prime}}=\alpha _m \right)}
\label{eqn:raw_combinedProb_xn}.
\end{equation}
After updating the convergence factor $\eta$, we compare it with the historical maximum convergence factor $\eta _{\max}$. If $\eta$ is less than $\eta_{\max}$ and their difference exceeds a constant $\epsilon$, the iteration is terminated; otherwise, the iteration continues. If $\eta$ exceeds $\eta_{\max}$, the value of $\eta_{\max}$ is updated to $\eta$, and the estimation of $\boldsymbol{x}$ is recalculated as
\begin{equation}
\left[ \hat{\boldsymbol{x}} \right] _{p^{\prime}}=\mathop {\mathrm{arg}\max} \limits_{\alpha _m\in \mathcal{X}}\,\,\bar{p}_{p^{\prime}}\left( \alpha _m \right). \label{eqn:x_estimation}
\end{equation}

\section{Simulation results and discussion}
In this section, we employ Monte Carlo simulations to evaluate the performance of the proposed GOCDM system. First, we compare the PAPR of GOCDM signals with that of OCDM signals. Subsequently, we evaluate the BER performance of GOCDM under two distinct doubly selective channels: a mobile underwater acoustic channel and a terrestrial mobile RF channel. The performance is benchmarked against traditional OCDM and OFDM signals.

\begin{figure}[!t]
	\centering
	\includegraphics[width=8.8cm]{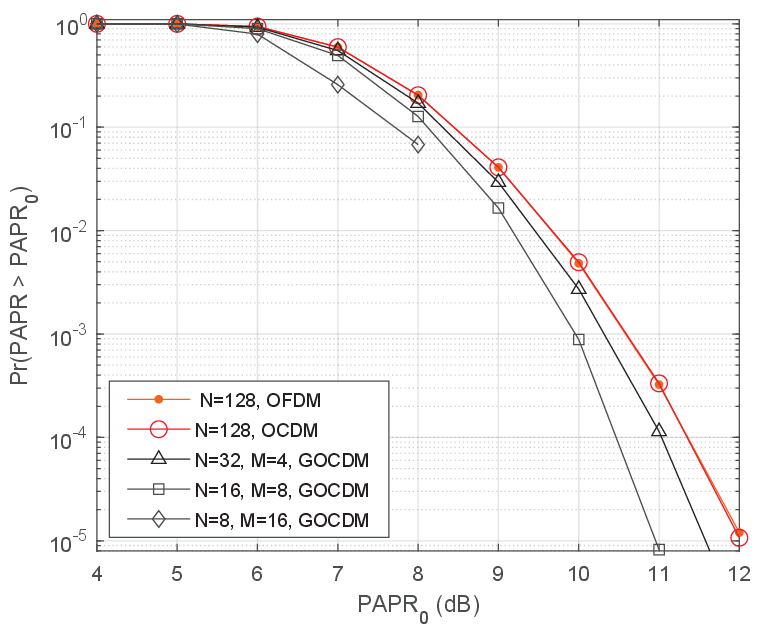}
	\caption{Comparison of the PAPR among GOCDM, OCDM and OFDM signals, where the number of orthogonal chirps in the OCDM symbols is N = 128, while GOCDM is configured with MN = 128. 4-QAM constellations are used in both systems.}\label{fig:PRAR}	
\end{figure}

Fig. \ref{fig:PRAR} depicts the comparison of the PAPR among the proposed GOCDM, conventional OCDM, and OFDM signals. The PAPR is calculated directly from the  baseband sequence of the block, denoted as $s_n$.  For a given symbol block, the PAPR is defined as $PAPR\,\,=\,\,\frac{P_{\max}}{P_{\mathrm{avg}}}$, where $P_{\max}$ is the maximum value of $\left| s_n \right|^2$ and $P_{\mathrm{avg}}$ is the average value of $\left| s_n \right|^2$ in the block. Both the GOCDM and OCDM baseband sequences have a length of 128, excluding the cyclic prefix, and employ a 4-QAM constellation.  Due to the randomness of the transmitted data bits, the PAPR of each signal block is also random. Let $PAPR_0$ be the PAPR threshold for the transmitted signal blocks, and $\mathrm{Pr}\left(PAPR>PAPR_0 \right)$ be the probability that the PAPR exceeds $PAPR_0$. For each waveform, we randomly generated $10^7$ symbol blocks to calculate the probability of event $PAPR>PAPR_0$. As demonstrated in the figure, in a statistical sense, the PAPR of GOCDM is significantly better than that of OCDM and OFDM. Additionally, for GOCDM signals, given a fixed symbol length (i.e., the product of $M$ and $N$ is constant), waveforms with smaller $N$ exhibit a better PAPR.

Next, we evaluate the bit error performance of GOCDM in time-frequency doubly selective channels. We first consider the application scenarios of GOCDM in underwater acoustic communications, where either the transmitter or the receiver, or both, are mounted on mobile underwater platforms. The configuration details for the transmitted signals are presented in Table \ref{table-simParas-UWA}. The relative velocity $V$ between the transmitter and receiver is 40 kilometers per hour. The speed of sound $C$ is 1500 meters per second. The channel consists of 9 physical paths, each with attenuation factors that follow a zero-mean complex Gaussian distribution. The attenuation factors of these paths are mutually independent. Table \ref{table-UWA-DelayPower} presents the time delays and the mean squared values of the power attenuation factors for each path. The Doppler shift for path $i$ is $v_i=v_{\max}\cos \left( \theta _i \right)$ where $v_{\max}$ represents the maximum Doppler shift, calculated from the relative velocity $V$, carrier frequency $f_c$, and sound speed $C$ as $v_{\max}=\frac{Vf_c}{C}$. Here, $\theta_i$ denotes the angle between the signal arrival direction for path $i$ and the relative motion direction of the transmitter and receiver, and $\theta_i$  is uniformly distributed in the interval $\left[ {-0.5\pi, 0.5\pi} \right]$. The angles $\theta_i$ are mutually independent across different paths. Based on the aforementioned parameters, the channel's time spread $S_t$ and Doppler Spread $S_f$ are calculated to be 14.7 milliseconds and 355.6 Hertz, respectively. The product $S_tS_f$ of the channel is 5.2, which is significantly greater than 1, indicating that the channel is an overspread channel and exhibits rapid time-variation characteristics. In the MP receivers for both GOCDM and OCDM, we set the damping factor $\Delta$ of the MP algorithm to 0.6. Each fractional Doppler shift path is equivalently simulated using 10  paths with integer Doppler shifts, i.e., $B_i=10$.

\begin{table}[tbp] \normalsize 
	\centering
	\caption{ Configuration Parameters for UWA Signals} 
	\begin{tabular}{l|l} 
		\hline
		Parameter & value \\
		\hline
		Carrier frequency $f_c$ & 24 kHz\\
		Bandwidth & 3.2 kHz \\
		No. of samples per block without CP & 128 \\
		Constellation & 4-QAM  \\
		Symbol block duration $T$ & 55 ms \\
		Guard interval & 15 ms \\
		\hline
	\end{tabular} 
	\label{table-simParas-UWA}
\end{table}

\begin{table}[tbp] \normalsize 
	\centering
	\caption{ Delay-power profile of the UWA Channel Paths} 
	\begin{tabular}{c|c} 
		\hline
		Path delay (mS)& Relative power (dB)\\
		\hline
		0 & 0 \\
		0.6 & -0.6 \\
		1.3 & -1\\
		2.2 &  -1.3\\
		6.9 &  -2.8\\
		7.5 & -4.2\\
		8.1 & -3.5\\
		13.1 & -6.2\\
		13.8 & -7.3\\
		14.7 & -8.1 \\
		\hline
	\end{tabular} 
	\label{table-UWA-DelayPower}
\end{table}

\begin{figure}[!t]
	\centering
	\includegraphics[width=8.8cm]{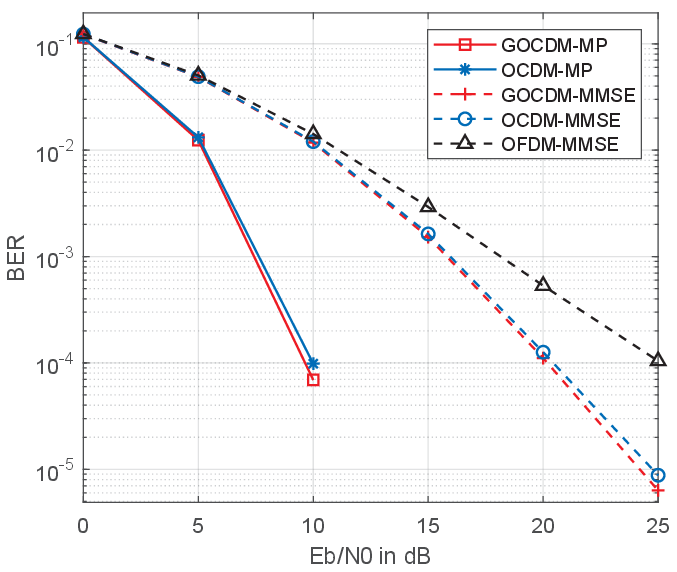}
	\caption{BER performance comparison for GOCDM, OCDM and OFDM in UWA mobile channels, where the number of orthogonal subchannels in OCDM (or OFDM) is set to be N = 128, while GOCDM is configured with MN = 128. 4-QAM constellations are used in all systems.}\label{fig:BER-UWA}	
\end{figure}

Fig. \ref{fig:BER-UWA} illustrates the comparison of bit error rate (BER) performance among GOCDM, OCDM, and OFDM under the above system settings. The horizontal axis represents the energy per bit to noise power spectral density ratio (Eb/N0), measured in decibels (dB). It should be noted that in all simulations presented in this paper, the calculation of the signal's energy per bit takes into account the overhead introduced by the cyclic prefix. As observed from the figure, whether using a receiver based on MMSE equalization or an iterative receiver based on the MP algorithm, the BER performance of GOCDM is slightly superior to that of OCDM. Additionally, it is evident that both GOCDM and OCDM significantly outperform OFDM when employing the MMSE receiver. This is because the energy of any data symbol in GOCDM (or OCDM)  is spread across all frequency subchannels of the channel, allowing the data symbol to experience the diversity of the subchannel frequency responses. In contrast, in OFDM, the energy of any given data symbol is allocated to a single subcarrier. Although the Doppler effect in the channel causes this symbol's energy to spread to adjacent subcarriers, the data symbol still does not fully experience the diversity of the channel's frequency response. 

Next, we evaluate the bit error performance of GOCDM in terrestrial RF doubly selective channels. We consider a vehicular mobile communication scenario where the relative velocity $V$ between the transmitter and receiver is 500 kilometers per hour. The carrier frequency, bandwidth, symbol duration, and other parameters of the transmitted signals (including GOCDM, OCDM, and OFDM signals) are detailed in Table \ref{table-RFsigParas}.  The delay-power profile of the channel paths is configured using the extended vehicular A model \cite{EVA-Model}, with specific parameters provided in Table \ref{table-radio-DelayPower}. Based on the relative velocity $V$ between the transmitter and receiver and the carrier frequency $f_c$, the maximum Doppler shift $v_{\max}$ of the signal is calculated to be 2315 Hz. Its normalized value is 0.0386 (with a normalization factor of $1/T$). The Doppler shifts of the channel paths are independent random variables, where the Doppler shift of path $i$ is given by $v_i = v_{\max}\cos(\theta_i)$), and $\theta_i$ is a uniformly distributed random variable in the interval $[-\pi/2, \pi/2]$. 
Since the normalized Doppler shifts of the paths are significantly less than 0.5, we approximate each fractional Doppler-shift path by 5 paths with integer Doppler shifts, i.e., $B_i = 5$. Similar to the example in UWA communication, the damping factor and the maximum allowed iterations for the MP algorithm are set to 0.6 and 20, respectively. It is noteworthy that, due to the near-light speed of electromagnetic wave propagation, the product of the delay spread and Doppler spread  $S_tS_f$ in a vehicular mobile communication channel is much less than 1, indicating that it is an underspread LTV channel.

Fig. 6 compares the BER performance of GOCDM, OCDM, and OFDM under the above underspread radio channel. All three systems have a symbol block length of 256 (ideal sampling points, excluding the cyclic prefix length). For the GOCDM system, the parameters $M$ and $N$ are set to 8 and 32, respectively. As depicted in the figure, the performance of GOCDM is almost the same to that of OCDM, regardless of whether MMSE equalizers or the MP algorithms are used for detection. Furthermore, both GOCDM and OCDM systems significantly outperform the OFDM system.

\begin{table}[t] \normalsize 
	\centering
	\caption{ Configuration parameters of the terrestrial RF Signals} 
	\begin{tabular}{l|l} 
		\hline
		Parameter & Value \\
		\hline
		Carrier frequency $f_c$ & 5 GHz\\
		Bandwidth & 15.360 MHz \\
		No. of samples per block without CP & 256 \\
		Constellation & 4-QAM  \\
		Symbol block duration $T$ & 19.27 $\mu$s \\
		Guard interval & 2.6 $\mu$s \\
		\hline
	\end{tabular} 
	\label{table-RFsigParas}
\end{table}

\begin{table}[t] \normalsize 
	\centering
	\caption{ Delay-Power Profile of the Extended Vehicular A Model} 
	\begin{tabular}{c|c} 
		\hline
		Path delay (nS)& Relative power (dB)\\
		\hline
		0 & 0 \\
		30 & -1.5 \\
		150 & -1.4\\
		310 &  -3.6\\
		370 &  -0.6\\
		710 & -9.1\\
		1090 & -7.0\\
		1730 & -12.0\\
		2510 & -16.9\\
		\hline
	\end{tabular} 
	\label{table-radio-DelayPower}
\end{table}

\begin{figure}[!t]
	\centering
	\includegraphics[width=8.8cm]{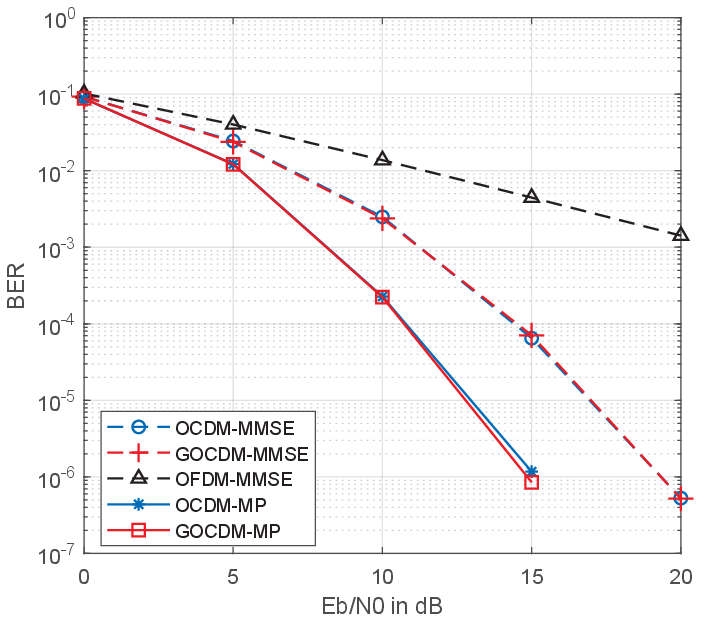}
	\caption{BER performance comparison for GOCDM, OCDM, and OFDM under EVA channel model with relative motion velocity of 500 kmph. The number of subcarriers in OFDM and the number of chirps in OCDM are both 256. For GOCDM, the parameters $M$ and $N$ are set to 8 and 32, respectively.}\label{fig:radio_BER}	
\end{figure}

\section{Conclusion}
This paper extends the conventional DFnT by introducing a unitary transform known as the GDFnT. Using this transform, data symbol vectors in the GF domain are converted into time-domain signal vectors, resulting in novel signal waveforms termed GOCDM. We derived the GF-domain channel matrix based on a time-frequency doubly selective channel characterized by multiple Doppler shifts of channel paths. By leveraging the sparsity of the GF-domain channel matrix, iterative detection for data symbols of GOCDM is performed using the message passing algorithm. Simulation results demonstrate that GOCDM offers significant advantages in terms of PAPR while achieving slightly better BER performance than OCDM under overspread DS channels, or nearly identical performance under underspread DS channels.

\appendices
\section{Proof of Lemma \ref{Lemma_1}}
With (\ref{eqn:P_i}), for any given $p,q = 0,1,\dots,MN-1$, it is easy to verify that 
\begin{equation}
[\mathbf{\Pi \Theta }_{M,N}^{H}]_{p,q}=\left[ \mathbf{\Theta }_{M,N}^{H} \right] _{\left< p-1 \right> _{MN},q},
\end{equation}
and
\begin{equation}
[\mathbf{\Theta }_{M,N}^{H}\mathbf{\Pi }]_{p,q}=\left[ \mathbf{\Theta }_{M,N}^{H} \right] _{p,\left< q+1 \right> _{MN}}.
\end{equation}
Therefore, we have 
\begin{align}
&\left[ \mathbf{\Pi \Theta }_{M,N}^{H}-\mathbf{\Theta }_{M,N}^{H}\mathbf{\Pi } \right] _{p,q} \nonumber \\
&=\left[ \mathbf{\Theta }_{M,N}^{H} \right] _{\left< p-1 \right> _{MN},q}-\left[ \mathbf{\Theta }_{M,N}^{H} \right] _{p,\left< q+1 \right> _{MN}} \nonumber
\\
&=\left[ \mathbf{\Theta }_{M,N} \right] _{q,\left< p-1 \right> _{MN}}^{*}-\left[ \mathbf{\Theta }_{M,N} \right] _{\left< q+1 \right> _{MN},p}^{*}, \label{eqn:proofLemma_1_3}
\end{align}
where the second equality in (\ref{eqn:proofLemma_1_3}) arises from the fact that $\mathbf{\Theta }_{M,N}^{H}$ is a unitary matrix.

Let's denote $p$ and $q$ as  $p\,\,=\,\,\lfloor \frac{p}{M} \rfloor M+\left< p \right> _M$ and $q=\,\,\lfloor \frac{q}{M} \rfloor M+\left< q \right> _M$, respectively. From (\ref{blockMat_theta}), we have 
\begin{align}
	\left[ \mathbf{\Theta }_{M,N}
	 \right] _{p,q}&=\left[ \mathbf{\Phi }_N \right] _{\lfloor \frac{p}{M} \rfloor ,\lfloor \frac{q}{M} \rfloor}\delta \left[ \left< p \right> _M-\left< q \right> _M \right]  \nonumber
	\\
	&=\left[ \mathbf{\Phi }_N \right] _{\lfloor \frac{p}{M} \rfloor ,\lfloor \frac{q}{M} \rfloor}\delta \left[ \left< p-q \right> _M \right]  \label{eqn:Theta_p_q}.
\end{align}

Using (\ref{eqn:Theta_p_q}), we can rewrite (\ref{eqn:proofLemma_1_3}) as
\begin{align}
&\left[ \mathbf{\Pi \Theta }_{M,N}^{H}-\mathbf{\Theta }_{M,N}^{H}\mathbf{\Pi } \right] _{p,q} \nonumber
\\
&=\left( \left[ \mathbf{\Phi }_N \right] _{\lfloor \frac{q}{M} \rfloor ,\lfloor \frac{\left< p-1 \right> _{MN}}{M} \rfloor}-\left[ \mathbf{\Phi }_N \right] _{\lfloor \frac{\left< q+1 \right> _{MN}}{M} \rfloor ,\lfloor \frac{p}{M} \rfloor} \right) ^* \nonumber
\\
&\quad \delta \left[ \left< q-p+1 \right> _M \right]. \label{eqn:proofLemma_1_4}
\end{align}
Now, let's discuss the value of the right-hand side of (\ref{eqn:proofLemma_1_4}) for different $p$ and $q$ as follows.

(1) The case $\left< q-p+1 \right> _M \ne 0$

$\left[ \mathbf{\Pi \Theta }_{M,N}^{H}-\mathbf{\Theta }_{M,N}^{H}\mathbf{\Pi } \right] _{p,q}$ =0, sine $\delta \left[ \left< q-p+1 \right> _M \right] =0$ when $\left< q-p+1 \right> _M \ne 0$ .

(2) The case $\left< q-p+1 \right> _M = 0$ 

Before initiating the discussion of this case, please note that the $N$-pint DFnT matrix $
\mathbf{\Phi }_N$ is a circulant matrix. Specifically, for any given $m$ and $n$, $m,n=0,1,\cdots ,N-1$, we have $\left[ \mathbf{\Phi }_N \right] _{m,n}=\left[ \mathbf{\Phi }_N \right] _{\left< m+1 \right> _N,\left< n+1 \right> _N}
$ \cite{OCDM_MP}. We then divide the case $\left< q-p+1 \right> _M = 0$ into four sub-cases, each of which is discussed as follows.

(2.1) The case $p=0$, $q=MN-1$ and $\left< q-p+1 \right> _M = 0$ 
\begin{align}
\left[ \mathbf{\Pi \Theta }_{M,N}^{H}-\mathbf{\Theta }_{M,N}^{H}\mathbf{\Pi } \right] _{p,q}&=\left( \left[ \mathbf{\Phi }_N \right] _{N-1,N-1}-\left[ \mathbf{\Phi }_N \right] _{0,0} \right) ^* \nonumber
\\
&=0
\end{align}

(2.2) The case $p=0$, $q \ne MN-1$ and $\left< q-p+1 \right> _M = 0$ 
\begin{align}
	\left[ \mathbf{\Pi \Theta }_{M,N}^{H}-\mathbf{\Theta }_{M,N}^{H}\mathbf{\Pi } \right] _{p,q} &=\left( \left[ \mathbf{\Phi }_N \right] _{\lfloor \frac{q}{M} \rfloor ,N-1}-\left[ \mathbf{\Phi }_N \right] _{\lfloor \frac{q}{M} \rfloor +1,0} \right) ^*
	 \nonumber \\
	&=0
\end{align}

(2.3) The case $p \ne 0$, $q=MN-1$ and $\left< q-p+1 \right> _M = 0$ 
\begin{align}
\left[ \mathbf{\Pi \Theta }_{M,N}^{H}-\mathbf{\Theta }_{M,N}^{H}\mathbf{\Pi } \right] _{p,q}&=\left( \left[ \mathbf{\Phi }_N \right] _{N-1,\lfloor \frac{p}{M} \rfloor -1}-\left[ \mathbf{\Phi }_N \right] _{0,\lfloor \frac{p}{M} \rfloor} \right) ^* \nonumber \\
	&=0
\end{align}

(2.4) The case $p \ne 0$, $q\ne MN-1$ and $\left< q-p+1 \right> _M = 0$ 
\begin{align}
	&\left[ \mathbf{\Pi \Theta }_{M,N}^{H}-\mathbf{\Theta }_{M,N}^{H}\mathbf{\Pi } \right] _{p,q} \nonumber
	\\
	&=\left( \left[ \mathbf{\Phi }_N \right] _{\lfloor \frac{q}{M} \rfloor ,\lfloor \frac{p-1}{M} \rfloor}-\left[ \mathbf{\Phi }_N \right] _{\lfloor \frac{q+1}{M} \rfloor ,\lfloor \frac{p}{M} \rfloor} \right) ^* \nonumber
	\\
	&=\begin{cases}
		\left[ \mathbf{\Phi }_N \right] _{_{\lfloor \frac{q}{M} \rfloor ,\lfloor \frac{p}{M} \rfloor -1}}^{*}-\left[ \mathbf{\Phi }_N \right] _{_{\lfloor \frac{q}{M} \rfloor +1,\lfloor \frac{p}{M} \rfloor}}^{*}, \mathrm{if}\,\,\left< p \right> _M=0  \nonumber \\
		\left[ \mathbf{\Phi }_N \right] _{_{\lfloor \frac{q}{M} \rfloor ,\lfloor \frac{p}{M} \rfloor}}^{*}-\left[ \mathbf{\Phi }_N \right] _{_{\lfloor \frac{q}{M} \rfloor ,\lfloor \frac{p}{M} \rfloor}}^{*}, \mathrm{if}\,\,\left< p \right> _M\ne 0 \nonumber\\
	\end{cases} \nonumber
	\\
   &=0
\end{align}
From the above, it is evident that for any $p$ and $q$, $p,q=0,1,\dots,MN-1$, the equality $
\left[ \mathbf{\Pi \Theta }_{M,N}^{H}-\mathbf{\Theta }_{M,N}^{H}\mathbf{\Pi } \right] _{p,q}
=0$ always holds. Therefore, we have $\mathbf{\Pi \Theta }_{M,N}^{H}=\mathbf{\Theta }_{M,N}^{H}\mathbf{\Pi }$, which completes the proof of Lemma \ref{Lemma_1}.



\end{document}